\documentclass{kluwer}

\usepackage{graphicx}

\newcommand{\Msun}{\, {\rm M}_{\odot}}
\newcommand{\Zsun}{\, {\rm Z}_{\odot}}
\newcommand{\AuA}[2]{ A{\rm\&}A\/, { #1, #2}}
\newcommand{\AuAS}[2]{  A{\rm \&}AS\/, { #1, #2}}
\newcommand{\ApJ}[2]{ ApJ\/, { #1, #2}}
\newcommand{\ApJS}[2]{ ApJS\/, {#1, #2}}

\newcommand{\AJ}[2]{ AJ\/, { #1, #2}}

\newcommand{\PASJ}[2]{ PASJ\/, { #1, #2}}

\newcommand{\MNRAS}[2]{ MNRAS\/, { #1, #2}}

\begin{document}
\begin{article}

\begin{opening}

\title{Chemodynamical gas flow cycles and their influence on the chemical evolution of dwarf irregular galaxies}

\author{A. \surname{Rieschick}}
\author{G. \surname{Hensler}} 

\institute{Institut f\"ur Theoretische Physik und Astrophysik, Universit\"at Kiel, Germany}

\runningtitle{Chemodynamical gas flow cycles in dIrrs}
\runningauthor{A.~Rieschick \& G.~Hensler}

\begin{abstract}
Here we investigate an exemplary chemodynamical evolutionary simulation of a dwarf irregular
galaxy.
By means of this model we demonstrate the existence of three gas mixing cycles: 
1) An inner {\it local cycle} mixing the metals
produced in stars locally, and 2) an outer {\it galactic cycle} on which hot gas is driven 
out of the galaxy by multiple supernovae type II and 
mixes on a short timescale with the available cold gas. 3) Only a small fraction of the metals leaves the 
galactic gravitational field and follows the {\it global cycle} with the intergalactic matter.
The large-scale mixing results in a temporary depletion of supernova ejected metals.
We will discuss this {\it delayed recycling} and its influence on the chemical evolution, 
especially on the nitrogen over oxygen ratio which is increased temporarily.
These results presented her are also relevant for less sophisticated analytical approaches and chemical 
evolutionary models of galaxies which have to parameterize the metal loss through outflow.
\end{abstract}
\end{opening}

\vspace{-2mm}
\section{Introduction}
\vspace{-2mm}
Many observations have been carried out to determine the chemical abundances in dwarf irregular galaxies (dIrrs) 
(e.g., \citeauthor{Garnett90} \citeyear{Garnett90};
\citeauthor{Pilyugin92} \citeyear{Pilyugin92}, \citeyear{Pilyugin93}). They show a wide scatter both in metal abundances and
in abundance ratios (e.g., \citeauthor{vZSH98} \citeyear{vZSH98}). 
In spite of this, the metal distribution inside some of the galaxies is rather homogeneous
(e.g., NGC~1569: \citeauthor{KS97} \citeyear{KS97}; I~Zw~18: \citeauthor{IT99} \citeyear{IT99}).

Metals produced in high-mass stars (HMSs) and intermediate-mass stars (IMSs) polute different gas phases.
Due to the diverse energy contents of supernovae (SNe) and planetary nebulae (PNe) their mass releases are stored in the hot 
intercloud medium (ICM) or the cold cloudy medium (CM), respectively. The CM-bound metals remain in the galactic body, while 
SN ejecta might leave the gravitational field of the galaxy. Some hydrodynamic simulations have shown 
on the one hand that galactic winds are able to expel the SN~II-ejected gas 
(e.g., \citeauthor{DYG90} \citeyear{DYG90}; \citeauthor{MLF99} \citeyear{MLF99}) while in contrast others
(e.g., \citeauthor{DB99} \citeyear{DB99}) allow the gas to be kept in the gravitational field of dIrrs if an extended gaseous halo exists.
\citeauthor{TT96} (\citeyear{TT96}) and \citeauthor{STT98} (\citeyear{STT98}) describe scenarios where the gas 
leaves the galaxy but rains back to other locations. Additionally, \citeauthor{KSRWR97} (\citeyear{KSRWR97}) have observed 
regions e.g. in NGC~5253 with larger Nitrogen and Oxygen abundances what can be explained by local self-enrichment.


\vspace{-2mm}
\section{Chemodynamical models}
\vspace{-2mm}
We have performed self-consistant chemodynamical evolution simulations. 
The code distinguishes between two gas phases necessary to investigate the separated N and, respectively, O contamination and
the abundance mixing processes. 
Three stellar mass ranges are treated to distinguish the different stellar properties for the element release.
Additionally to gravitation all relevant mass and energy exchange processes 
between stars and gas and the gas phases themselves are taken into account 
such as star formation, stellar mass loss and death, evaporation, condensation, drag forces, cloud-cloud collisions and radiative ionisation as
well as e.g. radiative cooling of the gas.
Details of these simulations can be found in \citeauthor{RH03} (\citeyear{RH03}). The chemodynamical treatment, 
its basic network of interaction processes and the numerical code are descriped in detail in the 
comprehensive paper by \citeauthor{SHT97} (\citeyear{SHT97}).
We assume that the galaxy starts from a
protogalactic gas cloud with a Plummer-Kuzmin density distribution (Satoh 1980) and a
baryonic mass of about $10^{10}~\Msun$ within a static dark matter halo according to 
\citeauthor{Burkert95} (\citeyear{Burkert95}) of $10^{11}~\Msun$.

\begin{figure}
  \centerline{\includegraphics[width=10cm]{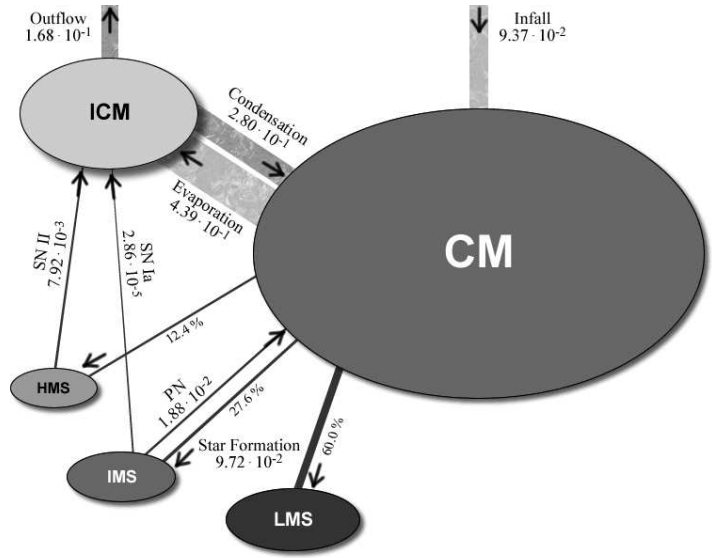}}
    \caption{\small
      Flow of matter between the components.
      The data are integrated over a region of radius $12.0~kpc$ and z-height $1.0~kpc$, i.e.
      the whole visible galaxy. The sizes of the ellipses are not scaled correctly according to the mass 
      of the components and have to be interpreted merely as a hint. The size of the mass flow pipes
      correspondes to the value of the mass exchange rate.}
    \label{Flow_Irr}
\end{figure}

\vspace{-3mm}
\section{Results}
\vspace{-2mm}
Here we focus on the different mass exchange processes acting within the galactic body. Fig.~\ref{Flow_Irr} shows all relevant
processes for the {\it irregular phase} of the dIrr model (see also \citeauthor{RH00} \citeyear{RH00} where a different
chemodynamical simulation was presented). On the
top right-hand side the "Infall" of CM from the gaseous outskirts into the galactic body is shown. In combination with evaporation and outflow of hot gas 
a large-scale cycle of mass flow is produced, because the ICM reaches the enveloping HI-halo 
where condensation rate exceeds evaporation. Therefore, the cycle is closed. 

It is important to notice that local self-enrichment by PNe occurs in the forming disk, because the time for evaporation and blow out
the complete CM (roughly $5\cdot10^{8}~\Msun$) exceeds $1~Gyr$. On the other hand, 
since the ICM phase contains much less mass (some $10^{6}~\Msun$), the complete ICM gas is condensated onto the CM
or blown out on a rather short timescale. 
Some of the SN II ejecta flow out directly, while another part leaves the galaxy not until it follows a detour through CM phase 
and evaporation process into ICM again. In total, only about $45 \%$ of the SN II metals are poluting the CM locally, while the rest is leaving 
the galactic body through ICM outflow. 

We divide the passage of gas and its metal content into three cycles:
\vspace{-5mm}
{\small
\begin{itemize}
\item {\bf {\it local cycle:}} The metals contained in PNe remain in the CM. About half of the SN II ejecta are
mixed into the local CM.
\vspace{-1mm}
\item {\bf {\it galactic cycle:}} Gas leaves the galactic body by high velocity outflow and is
mixed into the halo on a short timescale. The gas rains back into the galaxy with typical timescales of
$\approx 100~Myr$ up to $3~Gyr$.
\vspace{-1mm}
\item {\bf {\it global cycle:}} Less than five percent of the hot gas leaves the gravitational field poluting the intergalactic medium.
\end{itemize}
}

\vspace{-3mm}
\section{Discussion and conclusions}
\vspace{-2mm}
While PNe stay almost completely in the local SF regions, the chemodynamical dIrr simulations have shown that 
more than half of the SN II ejecta leave the galactic body before they reenter the inner, observable regions again. 
This {\it delayed recycling} has a distinct influence on the chemical evolution, especially on the ratio of nitrogen 
(N) to oxygen (O). 

Taking up-to-date stellar yields into account 
(\citeauthor{WW95} \citeyear{WW95}; \citeauthor{vdHG97} \citeyear{vdHG97}), 
O is almost completely processed in HMSs while N is produced both in HMSs but mostly in IMSs. 
In early evolutionary stages with lower metallicities only a small fraction of N originates from HMSs
while e.g. for $Z=0.1~\Zsun$ about $65~\%$ of a single stellar population's N yield is ejected by SNe II (secondary N production).
Comparing delayed recycling with instantaneous local mixing the N/O ratio is shifted to higher values up to about 1.0 dex.
This effect is due to the O deposition and
separated analytically from other dynamical and evolutionary influences, e.g. the different lifetime of HMSs and IMSs
which also influences the N/O ratio significantly (see e.g. \citeauthor{HEK00} \citeyear{HEK00}).
It has a particular importance for very metal-poor systems, but varies with the evolutionary phase as well as with global 
dynamical processes going on. Therefore chemodynamical models which consider all relevant dynamical and energetical processes 
can provide a fundamental insight into matter exchange and metal deposition in these sensitively balanced galaxies.

\vspace{-3.5mm}
\begin{acknowledgements}
\vspace{-2mm}
{\small
We gratefully acknowledge cooperation and
discussions with J.~K\"oppen, \linebreak Ch.~Theis and P.~Berczik. 
A.R. is supported by the {\it Deutsche Forschungsgemeinschaft} under grant no. He 1487/5-3 and 
He 1487/23-1.
}
\end{acknowledgements}

\end{article}


\vspace{-6.5mm}
\begin{thebibliography}{}
\vspace{-2mm}
\bibitem[\protect\citeauthoryear{Burkert}{1995}]{Burkert95} Burkert, A. 1995, \ApJ{447}{L25}
\bibitem[\protect\citeauthoryear{D'Ercole \& Brighenti}{1999}]{DB99} D'Ercole, A., Brighenti, F. 1999, \MNRAS{309}{941}
\bibitem[\protect\citeauthoryear{De Young \& Gallagher}{1990}]{DYG90} De Young, D.~S., Gallagher, J.~S. 1990, \ApJ{356}{L15}
\bibitem[\protect\citeauthoryear{Garnett}{1990}]{Garnett90} Garnett, D.~R. 1990, \ApJ{363}{142}
\bibitem[\protect\citeauthoryear{Henry et al.}{2000}]{HEK00} Henry, R.~B.~C., Edmunds, M.~G., K\"oppen, J. 2000, \ApJ{541}{660}
\bibitem[\protect\citeauthoryear{Izotov \& Thuan}{1999}]{IT99} Izotov, Y.I, Thuan, T.~X. 1999, in Proc. XVIII. Recontre de Moriond, 
  Les Arcs, {\it Dwarf Galaxies and Cosmology}, eds. T.~X.~Thuan, et al. (Gif-sur-Yvettes: Fronti\'eres), 223
\bibitem[\protect\citeauthoryear{Kobulnicky \& Skillman}{1997}]{KS97} Kobulnicky, H.~A., Skillman, E.~D. 1997, \ApJ{489}{636}
\bibitem[\protect\citeauthoryear{Kobulnicky \& Skillman}{1997}]{KSRWR97} Kobulnicky, H.~A., Skillman, E.~D.,
  Roy, J.-R., Walsh, J.~R., Rosa, M.~R. 1997, \ApJ{477}{679}
\bibitem[\protect\citeauthoryear{Mac Low \& Ferrara}{1999}]{MLF99} Mac Low, M.~M., Ferrara, A. 1999, \ApJ{513}{142}
\bibitem[\protect\citeauthoryear{Pilyugin}{1992}]{Pilyugin92} Pilyugin, L.~S. 1992, \AuA{260}{58}
\bibitem[\protect\citeauthoryear{Pilyugin}{1993}]{Pilyugin93} Pilyugin, L.~S. 1993, \AuA{277}{42}
\bibitem[\protect\citeauthoryear{Rieschick \& Hensler}{2000}]{RH00} Rieschick, A., Hensler, G. 2000, {\it Cosmic Evolution and Galaxy Formation: Structure, 
  Interactions and Feedback}, ASP Conf. Ser. eds. J.~Franco, E.~Terlevich, O.~L\`opez~Cruz \& I.~Aretxaga, 
  Astron.~Soc.~Pac., San Francisco
\bibitem[\protect\citeauthoryear{Rieschick \& Hensler}{2003}]{RH03} Rieschick, A., Hensler, G. 2003, A{\rm\&}A\/, to be submitted
\bibitem[\protect\citeauthoryear{Samland et al.}{1997}]{SHT97} Samland, M., Hensler, G., Theis, Ch. 1997, \ApJ{476}{544}
\bibitem[\protect\citeauthoryear{Satoh}{1980}]{Satoh80} Satoh, C. 1980, \PASJ{32}{41}
\bibitem[\protect\citeauthoryear{Silich \& Tenorio-Tagle}{1998}]{STT98} Silich, S.~A., Tenorio-Tagle, G. 1998, \MNRAS{299}{249}
\bibitem[\protect\citeauthoryear{Tenorio-Tagle}{1996}]{TT96} Tenorio-Tagle, G. 1996, \AJ{111}{1641}
\bibitem[\protect\citeauthoryear{van den Hoek \& Groenewegen}{1997}]{vdHG97} van den Hoek, L.~B., Groenewegen, M.~A.~T. 1997, \AuAS{123}{305}
\bibitem[\protect\citeauthoryear{van Zee et al.}{1998}]{vZSH98} van Zee, L., Salzer, J.~J., Haynes, M.~P. 1998, \ApJ{497}{L1}
\bibitem[\protect\citeauthoryear{Woosley \& Weaver}{1995}]{WW95} Woosley, S.~E., Weaver, T.~A. 1995, \ApJS{101}{181}
\end{thebibliography}
\end{document}